\documentclass[aps,prl,twocolumn,superscriptaddress,floatfix,10pt,balancelastpage]{revtex4-1}
\pdfoutput = 1
\usepackage{graphicx}
\usepackage{pdfpages}
\usepackage{hyperref}
\hypersetup{urlcolor=blue, linkcolor=blue, citecolor=blue, colorlinks=true}

\begin{document}
\title{Multiple Quantum Phases in Graphene with Enhanced Spin-Orbit Coupling:\\ From the Quantum Spin Hall Regime to the Spin Hall Effect and a Robust Metallic State}
\author{Alessandro Cresti}
\affiliation{Univ. Grenoble Alpes, IMEP-LAHC, F-38016 Grenoble, France}
\affiliation{CNRS, IMEP-LAHC, F-38016 Grenoble, France}
\author{Dinh Van Tuan}
\affiliation{ICN2 - Institut Catala de Nanociencia i Nanotecnologia, Campus UAB, 08193 Bellaterra (Barcelona), Spain}
\affiliation{Department of Physics, Universitat Aut\'{o}noma de Barcelona, Campus UAB, 08193 Bellaterra, Spain}
\author{David Soriano}
\affiliation{ICN2 - Institut Catala de Nanociencia i Nanotecnologia, Campus UAB, 08193 Bellaterra (Barcelona), Spain}
\author{Aron W. Cummings}
\affiliation{ICN2 - Institut Catala de Nanociencia i Nanotecnologia, Campus UAB, 08193 Bellaterra (Barcelona), Spain}
\author{Stephan Roche}
\affiliation{ICN2 - Institut Catala de Nanociencia i Nanotecnologia, Campus UAB, 08193 Bellaterra (Barcelona), Spain}
\affiliation{ICREA - Institucio Catalana de Recerca i Estudis Avan\c cats, 08010 Barcelona, Spain}

\begin{abstract}
We report an intriguing transition from the quantum spin Hall phase to the spin Hall effect upon segregation of thallium adatoms adsorbed onto a graphene surface. Landauer-B\"uttiker and Kubo-Greenwood simulations are used to access both edge and bulk transport physics in disordered thallium-functionalized graphene systems of realistic sizes. Our findings not only quantify the detrimental effects of adatom clustering in the formation of the topological state, but also provide evidence for the emergence of spin accumulation at opposite sample edges driven by spin-dependent scattering induced by thallium islands, which eventually results in a minimum bulk conductivity $\sim 4e^{2}/h$, insensitive to localization effects. 
\end{abstract}

\maketitle

{\it Introduction.}-- 
In 2005, Kane and Mele predicted the existence of the quantum spin Hall effect (QSHE) in graphene due to intrinsic spin-orbit coupling (SOC) \cite{KAN_PRL95,KAN_PRL95b}.
Within the QSHE, the presence of spin-orbit coupling, which can be understood as a momentum-dependent magnetic field coupling to the spin of the electron, results in the formation of chiral edge channels for spin up and spin down electron populations. The observation of the QSHE is, however, prohibited in clean graphene owing to vanishingly small intrinsic spin-orbit coupling on the order of $\mu$eV \cite{YAO_PRB75}, but demonstrated in strong SOC materials (such as CdTe/HgTe/CdTe quantum wells or bismuth selenide and telluride alloys), giving rise to the new exciting field of topological insulators \cite{BER_PRL96,BER_SCI314,HAS_RMP82,QI_RMP83}.
Recent proposals to induce a topological phase in graphene include functionalization with heavy adatoms \cite{WEE_PRX1,JIA_PRL109}, covalent functionalization of the edges \cite{AUT_PRB87}, proximity effect with topological insulators \cite{JIN_PRB87,LIU_PRB87,KOU_NL13}, or intercalation and functionalization with 5$d$ transition metals \cite{HU_PRL109,LI_PRB87}.
In particular, the seminal theoretical study \cite{WEE_PRX1} by Weeks and co-workers has revealed that graphene endowed with a modest coverage of heavy adatoms (such as indium and thallium) could exhibit a substantial band gap and QSH fingerprints (detectable in transport or spectroscopic measurements). For instance, one signature of such a topological state would be a robust quantized two-terminal conductance ($2e^{2}/h$), with an adatom density-dependent conductance plateau extending inside the bulk gap induced by SOC \cite{WEE_PRX1,QIA_PRB82,QIA_PRL107}. To date, such a prediction lacks experimental confirmation \cite{JIA_arXiv}, despite some recent results on indium-functionalized graphene that have shown a surprising reduction of the Dirac point resistance with increasing indium density \cite{COR_ACR46}. On the other hand, it is known that adatoms deposited on two-dimensional materials inevitably segregate, forming islands rather than a homogeneous distribution \cite{SUT_SS17}, which significantly affects doping \cite{PI_RPB80,SAN_PRB84}, transport \cite{MCC_PRB81,ALE_PRB86,EEL_PRL110,KAT_PRB79} and optical \cite{YUA_PRB84,YUA_PRB90} features. The impact of adatom clustering on the formation of the QSHE remains, however, to be clarified.

Another fundamental mechanism induced by spin-orbit coupling in metals is the spin Hall effect (SHE), which is defined by the flow of bulk up-spin and down-spin currents in opposite directions (induced by spin-orbit scattering) without net charge current but resulting in spin accumulation at the lateral sample boundaries \cite{Vignale_2010,MAE_2011}. A recent experiment shows that the presence of physisorbed and clustered transition metals on graphene can give rise to an unexpectedly giant SHE \cite{BAL_NC5}, which has been assigned to resonant skew scattering induced by adatoms \cite{Ferreira_2014}. 

This Letter shows that by functionalizing graphene with heavy adatoms (such as thallium) it is possible to observe multiple quantum regimes driven by SOC including the QSHE, the SHE, and an unconventional bulk metallic state, depending on the characteristics of adatom clustering (adatom density, island size). The robustness of the QSHE is guaranteed for large enough and homogenous adatom distribution onto graphene, and for moderate adatom clustering into small islands. The topological state and chiral edge currents disappear when adatom segregation inhibits a bulk band gap, but a SHE-like phase further develops as evidenced by the formation chiral currents in the bulk and spin accumulation at the sample edges. Further increasing the island size eventually jeopardizes the SHE, but an unconventional transport regime remains owing to local chiral spin currents around thallium islands, characterized by a finite value of the Dirac point conductivity $\sim 4e^{2}/h$ and an absence of the insulating regime. Remarkably, the formation of locally chiral spin currents, rotating clockwise or anticlockwise around the islands depending on their spin polarization, is the bottom line of the multiple quantum phases occurring in such graphene materials with enhanced intrinsic spin-orbit coupling.

\begin{figure}[tb]
  \resizebox{8cm}{!}{\includegraphics{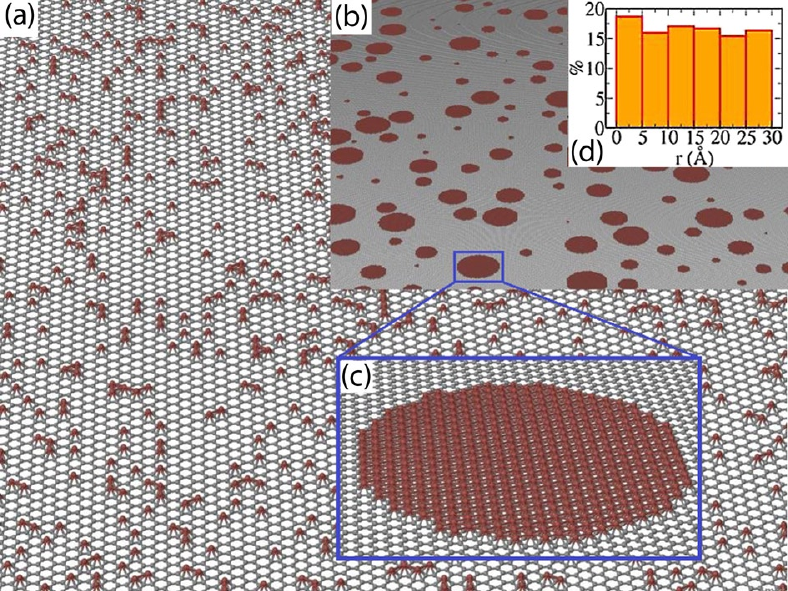}}
  \caption{(a) Ball-and-stick model of a graphene substrate with randomly adsorbed thallium atoms (concentration is  $15\%$). (b) Same as (a) but with adatoms clustered in islands with a radius distribution varying up to 3 nm (histogram shown in (d)). (c) Zoom-in of a typical thallium island. All thallium atoms are positioned in the hollow position and equally connected to the 6 carbon atoms forming the hexagon underneath (following \cite{WEE_PRX1}).}
 \label{fig1}
\end{figure}

{\it Model and methods.}--
When a thallium atom is grafted onto graphene, it sits in the middle of a hexagonal plaquette of carbon atoms, above the surface, see Fig\ref{fig1}. As shown in \cite{WEE_PRX1}, the adatom effect can be described in the effective $\pi$-$\pi$* tight-binding model including SOC. In the presence of adatoms randomly distributed over a set $\mathcal{R}$ of plaquettes, the Hamiltonian \cite{KAN_PRL95,KAN_PRL95b} reads
\begin{eqnarray}
\hat{H}=&-&\gamma\sum_{\langle ij\rangle }c_i^\dag c_j+\frac{2i}{\sqrt{3}}\lambda\sum_{\langle\langle ij\rangle\rangle \in \mathcal{R}}c_i^\dag\vec{s}\cdot(\vec{d}_{kj}\times\vec{d}_{ik})c_j\nonumber\\
&-&\mu\sum_{i\in \mathcal{R}}c_i^\dag c_i+\sum_{i}V_ic_i^\dag c_i \ ,
\label{Hamil}
\end{eqnarray}
where $c_i=(c_{i\downarrow},c_{i\uparrow})$ is the pair of annihilation operators for electrons with spin down and spin up on the $i$th carbon atom, and $c_i^\dag$ is the corresponding pair of creation operators.
The first term in Eq.(\ref{Hamil}) is the nearest neighbor hopping with coupling energy $\gamma=2.7$ eV. The second term is a next nearest neighbor hopping that represents the intrinsic SOC induced by the adatoms, with $\vec{d}_{kj}$ and $\vec{d_{ik}}$ the unit vectors along the two bonds connecting second neighbors and $\vec{s}$ the spin Pauli matrices.
The SOC is set to $\lambda=0.02\gamma$ from \textit{ab initio} simulations \cite{WEE_PRX1}. The third term describes the potential energy induced by charge transfer between adatoms and graphene. The last term  represents  the long-range interaction of graphene and impurities in the substrate  $V_i=\sum_{j=1}^{N}\epsilon_j\exp[-({\bf r}_i-{\bf R}_j)^2/(2\xi^2)]$ \cite{ORT_EPL94}, where $\xi=0.426$ nm is the effective range and the sum runs over $N$ impurity centers with random positions ${\bf R}_j$ and strength of the potential $\epsilon_j$ randomly chosen within $\left[-\Delta,\Delta\right]$. The Hamiltonian does not consider the effect of structural relaxation in the case of clustered adatoms. This is not expected to alter our conclusions.

For studying electronic transport in thallium-functionalized ribbons, we consider a standard two-terminal configuration with highly doped contacts, and compute the conductance using nonequilibrium Green's function formalism \cite{CRE_EJPB53}. The doping is mimicked by an appropriate potential energy $V$ on the source and drain. The differential conductance is obtained by the Landauer-B\"uttiker formula
\begin{equation}
	G(E) = (e^2/h) {\rm Tr} [G^R(E)\Gamma^{\rm (S)}G^A(E)\Gamma^{\rm (D)}] \ ,
\end{equation}
where $G^{R/A}$ are the retarded and advanced Green's functions and $\Gamma^{\rm (S/D)}$ are the rate operators for the source and drain contacts. This approach also gives access to the spin-resolved local spectral currents. Note that the absence of spin-mixing terms in the Hamiltonian (\ref{Hamil}) means that $S_z$ is a good quantum number and spin currents are well defined, since there is no electron flux between states with opposite spin \cite{note}.

We also study quantum transport in two-dimensional functionalized graphene by means of the Kubo approach \cite{FOA_2014,ROC_SSC152,YUA_PRB82}. The scaling properties of the conductivity can be followed through the dynamics of electronic wave packets using $\sigma(E,t)=e^{2}\rho(E)D(E,t)$, where $D(E,t)=\Delta X^{2}(E,t)/t$ is the diffusion coefficient, $\rho(E)$ is the density of states, and $\Delta X^{2}(E,t) = Tr\left[\delta(E-\hat{H})\left|\hat{X}(t)-\hat{X}(0)\right|^2\right]/Tr\left[\delta(E-\hat{H})\right]$ [with $\hat{X}(t)$ the position operator in Heisenberg representation] is the energy- and time-dependent mean quadratic displacement of the wave packet \cite{FOA_2014}.
Calculations are performed on systems containing more than $3.5\times10^6$ carbon atoms, corresponding to sizes larger than $300\times 300~{\rm nm}^{2}$. Such a size guarantees self-averaging, so that our results are weakly dependent on the specific spatial distribution of adatoms or clusters of adatoms. 

\begin{figure}[tb]
  \resizebox{8cm}{!}{\includegraphics{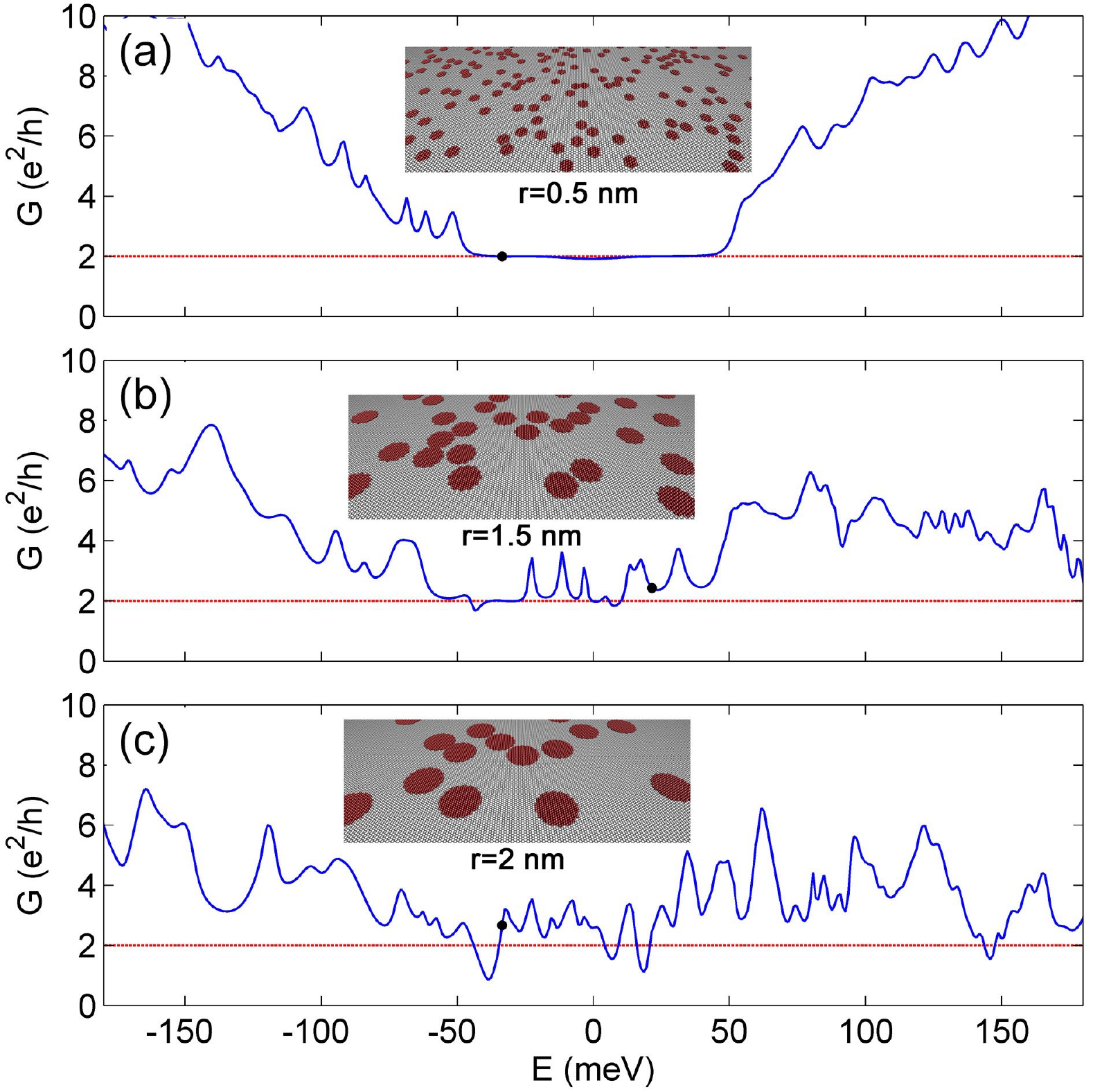}}
  \caption{Differential conductance for an armchair ribbon of width $W$=50 nm with a concentration $n$=15\% of thallium adatoms segregated into islands with radius $r=0.5$ nm (a), $r=1.5$ nm (b), and $r= 2$ nm (c), over a section with length $L$=50 nm. The potential energy on the contacts is set to $V=-2.5$ eV. The dotted lines indicate the conductance value $2e^2/h$.}
  \label{fig2}
\end{figure}

\begin{figure}[tb]
  \resizebox{8cm}{!}{\includegraphics{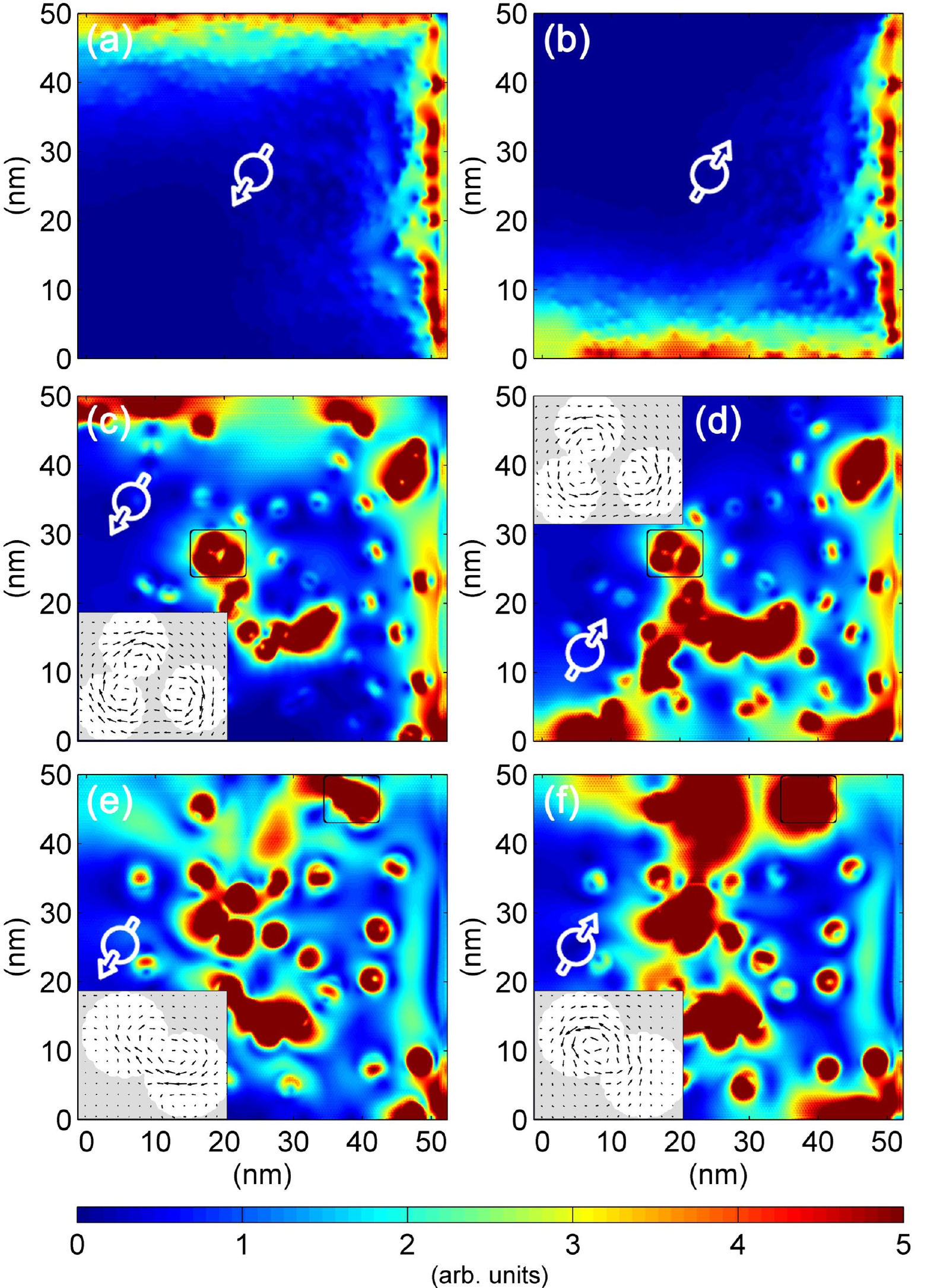}}
  \caption{Spin-resolved spectral current distribution for $r=0.5$ nm and $E=-33.5$ meV (a),(b); $r=1.5$ nm and $E=21.5$ meV (c),(d); and $r=2$ nm and $E=-33.5$ meV (e),(f). The corresponding energies and conductance are indicated by black dots in Fig.\ref{fig2}. The insets in panels (c-f) illustrate the local average current distribution in the regions indicated by the squares.} 
  \label{fig3}
\end{figure}

{\it Stability of the QSHE upon adatom deposition and clustering.}--
We start by considering an armchair ribbon of width $W$=50 nm functionalized with a concentration $n$=15\% of thallium adatoms segregated into islands with radius $r=0.5$ nm over a length $L$=50 nm [inset of Fig.\ref{fig2}(a)]. The differential conductance clearly shows a $2e^{2}/h$ plateau, which is the fingerprint of the QSHE [continuous line in Fig.\ref{fig2}(a)]. Because of the small size of the clusters, this behavior is analogous to that observed for nonsegregated adatoms \cite{WEE_PRX1}.
Note that the energy axis has been shifted by $3n\mu=$121.5 meV to compensate the charge neutrality point variation induced by the concentration ($\sim 3n$) of carbon atoms that undergo a charge transfer doping effect.
The width of the plateau is about 100 meV and approximately corresponds to the topological gap induced by thallium functionalization, i.e. $6\sqrt{3}\lambda_{\rm eff}\approx$84.2 meV, where the effective SOC is $\lambda_{\rm eff}=n \lambda \approx$8.1 meV \cite{SHE_PRB85}. This formula sheds light on the interplay between SOC and adatom density in determining the rise of a topological phase. For example, adatoms with a smaller SOC, such as indium, would require a higher density to obtain the same effect as thallium.
A closer inspection of the conductance shows that the plateau region is not perfectly flat, but varies within the range [1.92,2.02] $e^{2}/h$. This indicates that the separation between spin polarized chiral edge channels is not fully complete, partly because of existing randomness in the adatom distribution. 
Figures \ref{fig3}(a) and \ref{fig3}(b) show the spin-resolved local spectral current for electrons injected from the right contact at energy $E\approx-33.5$ meV, indicated by the dot in Fig.\ref{fig2}(a). 
We observe that the electrons injected from the source ($x>50$ nm) are rapidly and completely deviated along the top edge for spin down, see  Fig.\ref{fig3}(a), and along the lower edge for spin up, see Fig.\ref{fig3}(b). The width of the polarized edge channels in armchair ribbons does not depend on the energy but only on the SOC \cite{PRA_JCE12} as $a\gamma/(2\sqrt{3}\lambda_{\rm eff})\approx$13.5 nm, where $a$ is the carbon interatomic distance. The separation between the right-to-left and left-to-right moving channels, which is opposite for different spin polarizations, is at the origin of the QSHE. Note that no current flows through the bulk, where a topological gap is present.

This picture is markedly modified when adatoms segregate into larger islands. Figure \ref{fig2}(b) shows the differential conductance when the islands have a radius $r=1.5$ nm (the adatom concentration is kept at $n=15$\%). While a residual signature of the conductance plateau remains for certain energies, it is clear that quantization is almost completely lost despite the short intercluster distance. This indicates that adatom segregation has a detrimental effect on the formation of a topological phase in graphene. Considering that adatom clustering is unavoidable at room temperature, our findings provide an explanation for the absence of experimental confirmation of the existence of the QSHE in chemically functionalized graphene. 

An illuminating insight into the effect of segregation is further provided by the spin-resolved spectral current distribution reported in Figs.\ref{fig3}(c) and \ref{fig3}(d) for the energy $E=21.5$ meV indicated by the black dot in Fig.\ref{fig2}(b), where the conductivity is about 2.4 $e^2/h$, i.e., above the plateau. In the absence of a topological bulk band gap, the spectral current flows within the bulk, excluding the formation of the QSHE.  

However, a residual spin polarization still develops at the sample edges, as a result of spin-dependent bulk scattering of the electrons on thallium islands, which resembles the SHE mechanism. To corroborate this picture, the insets of Figs.\ref{fig3}(c) and \ref{fig3}(d) show the local current direction and intensity around the high current region indicated by a square, where three islands are present. It turns out that the current direction is clockwise for spin-down electrons and counterclockwise for spin-up electrons. This suggests that the scattering depends on the spin polarization, which drives electrons with opposite spins to be deflected at opposite edges, resulting in a spin accumulation. This mechanism is usually known as the spin Hall effect \cite{Vignale_2010,MAE_2011}.

When further increasing the island radius to $r=2$ nm or beyond, the conductance value globally shifts well above $2e^2/h$, in line with a deeper penetration of currents through the bulk [Fig.\ref{fig2}(c)]. This picture is illustrated by the local distribution of spectral currents in Fig.\ref{fig3}(e) and\ref{fig3}(f), where  edge spin accumulation is absent  or highly degraded \cite{SM}, thus impeding the observation of SHE.  Note, however, that the electron flow around the islands is still spin dependent, as shown in the insets of Figs.\ref{fig3}(e) and \ref{fig3}(f). 

We can anticipate the expected impact of the island size and the adatom density. For a given adatom density, a larger island radius translates into less 	numerous and more isolated clusters, with an even more pronounced suppression of the QSHE and the SHE. Increasing the adatom density would have the opposite effect, as the thallium coverage would be more thorough. However, even in the absence of a SHE, the clustering of thallium adatoms still produces a remarkable bulk transport fingerprint of the spin-orbit coupling in two-dimensional graphene, as discussed below. This will be shown to be related to the residual local chirality of bulk currents seen in Figs.\ref{fig3}(c)-(f).

\begin{figure}[tb]
  \resizebox{8cm}{!}{\includegraphics{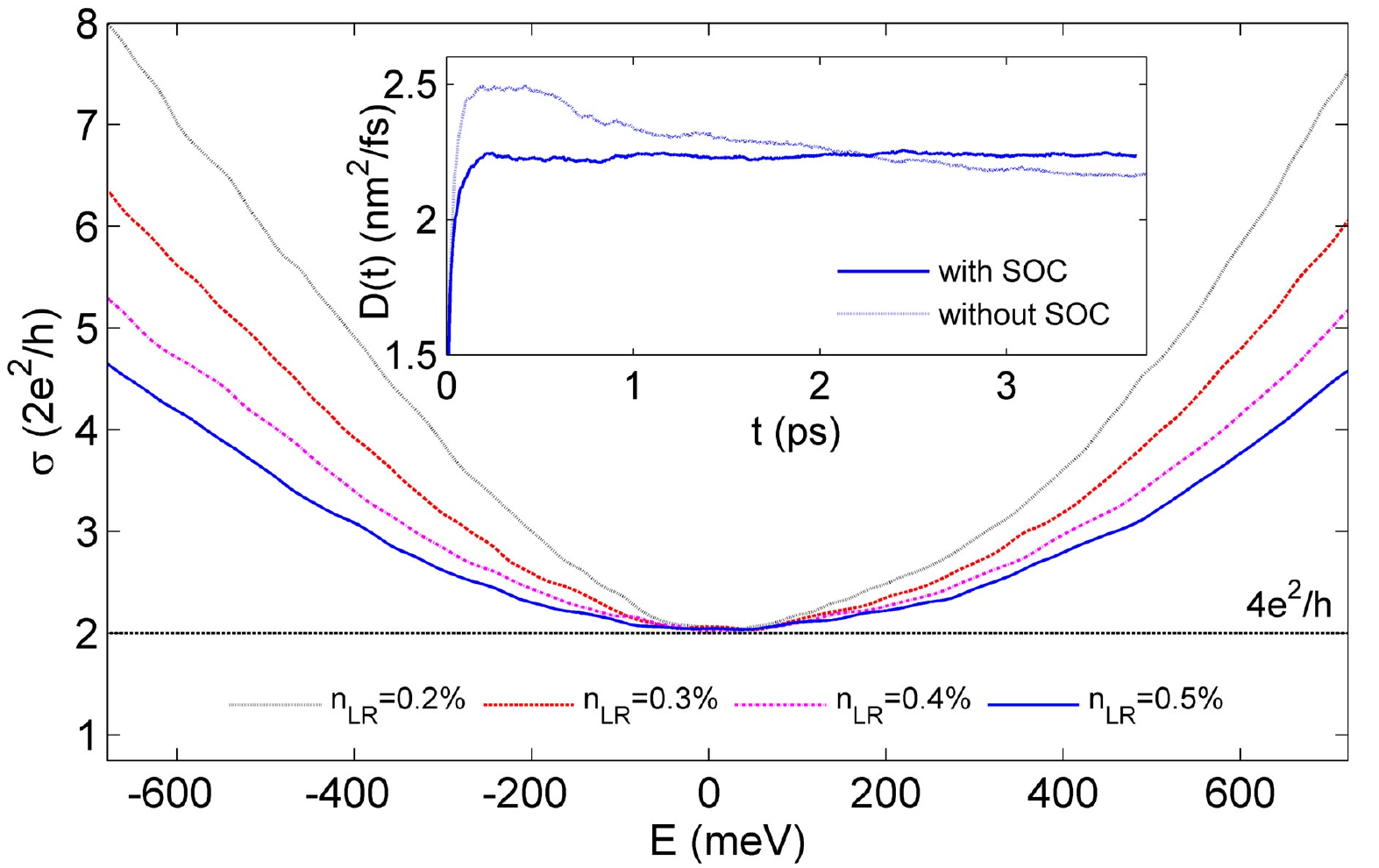}}
  \caption{Main frame: Kubo conductivity for thallium clustering and superimposed density ($n_{\rm LR}$) of long-range impurities. Inset: diffusion coefficient for wave packet with energy $E\approx 0$, for the case $n_{\rm LR}=0.5\%$, with (solid blue line) and without SOC.}
 \label{fig4}
\end{figure}

{\it Robust metallic state, minimum conductivity and chiral bulk currents.}--
We investigate the intrinsic bulk conductivity of thallium-functionalized graphene by computing the Kubo-Greenwood conductivity. We keep the same large thallium density (of about 15\%), but with thallium clusters size distribution given in Fig.\ref{fig1}(d), and consider a superimposed distribution of long-range impurities to mimic additional sources of disorder (such as charged impurities). In Fig.\ref{fig4} (main panel), the Kubo conductivity is shown for various densities ($n_{\rm LR}=0.2\%-0.5\%$) of long-range impurities with $\Delta=2.7$ eV. A striking feature is seen in the versatile impact of additional disorder on the  energy-dependent transport characteristics. A plateau is formed near the Dirac point, where the conductivity reaches a minimum value, regardless of the superimposed disorder potential. In contrast, a more conventional scaling behavior $\sigma\sim 1/n_{\rm LR}$ is obtained for higher energies, following the semiclassical Fermi golden rule. 
The minimum conductivity $\sigma_{\rm min}\sim 4e^{2}/h$ is reminiscent of the case of clean graphene deposited on oxide substrates and sensitive to electron-hole puddles \cite{DAS_RMP83}. However, the role of spin-orbit interaction here is critical for the formation of an anomalously robust metallic state. 
This is further shown in Fig.\ref{fig4} (inset), where the time dependence of the diffusion coefficient at energy $E\approx 0$ is reported for $n_{\rm LR}=0.5\%$, in the presence of thallium islands with and without SOC. The absence of SOC irremediably produces an insulating state, as evidenced by the decay of the diffusion coefficient. Conversely, once the SOC is switched on, the diffusivity saturates to its semiclassical values, showing no sign of localization.  The origin of such a suppression of localization by SOC is actually disconnected from weak antilocalization  \cite{MCC_PRL97,MCC_PRL108,RYC_EPL79,ZHA_PRL102,ORT_EPL94}, and instead is rooted in the formation of locally chiral bulk currents in the region of the thallium islands [see in Figs.(3)(e)-(f)].

Indeed, the existence of local chirality is expected to considerably reduce backscattering in the region of each impurity island, thus restricting localization phenomena \cite{SM}. It should be noted that since the spatial separation between chiral currents is not complete, scattering still  occurs in between islands preserving the diffusive nature of transport in the system. The observed partial chirality is obviously lost in the absence of SOC, as demonstrated by the time-dependent decay of the diffusion coefficient, consistent with the presence of localization effects [inset in Fig.\ref{fig4}]. This result is reminiscent of the prediction of a quantum critical point for Dirac point physics in the presence of random magnetic fields and long-range disorder \cite{OST_PRL98}.
Finally, a possible analogy might be established with the recently discovered bulk topological currents related to a valley Hall effect in graphene superlattices \cite{GOR_SCI346}.

{\it Conclusion.}-- The robustness of the QSHE in chemically functionalized graphene with enhanced SOC has been studied as a function of adatom clustering. A sufficiently large inhomogeneity of adatom segregation eventually quenches the topological bulk gap. In an intermediate regime, the formation of the SHE is driven by spin-dependent scattering effects on thallium adatom islands (a fact which supports the recent observation of SHE in CVD graphene in the presence of gold or copper adatoms\cite{BAL_NC5}). Ultimately, the existence of an anomalously robust metallic state close to the Dirac point is facilitated by local chiral currents concentrated around thallium islands. These findings might guide future experiments on the exploration of multiple quantum phases in chemically functionalized graphene \cite{COR_ACR46,KOU_NL13}. 

\textit{Acknowledgements.}-- S.R. acknowledges the Spanish Ministry of Economy and Competitiveness for funding (MAT2012-33911), the Severo Ochoa Program (MINECO SEV-2013-0295), and the Secretaria de Universidades e Investigaci\'on del Departamento de Econom\'ia y Conocimiento de la Generalidad de Catalu\~{n}a. The research leading to these results has received funding from the European Union Seventh Framework Programme under Grant Agreement No. 604391 Graphene Flagship.

\clearpage 

\section*{Supplementary material}

Supplemental Material

# Multiple Quantum Phases in Graphene with Enhanced Spin-Orbit Coupling: From the Quantum Spin Hall Regime to the Spin Hall Effect and a Robust Metallic State

*A. Cresti, D. Van Tuan, D. Soriano, A. W. Cummings, and S. Roche*

In this Supplemental Material, we first briefly comment the adopted model for spin-orbit coupling. Then, we show some simulations of the spin accumulation for the systems considered in the main text. Finally, we present a comprehensive cartoon to explain the mechanism behind the transition from the quantum spin Hall effect to the regimes with bulk chiral currents.

## Model for spin-orbit coupling effects due to thallium adatoms

The tight-binding Hamiltonian (1) (taken from [1]) used for describing the effect of thallium adatoms on graphene neglects the contribution of a Rashba-type spin-orbit coupling term. For the considered adatom densities and homogeneous distribution, its incorporation into (1) with $\lambda_R$ value as large as 58 meV is found to have a vanishing contribution in the formation of the QSH phase. This also agrees with the general criterion proposed by Kane and Mele [2], which states that although the Rashba term violates $S_z$ conservation, as long as $\lambda_R < 2\sqrt{3}\,\lambda_{SO}$, there is a finite region of the phase diagram that is adiabatically connected to the QSH phase at $\lambda_R = 0$.

For the situation of adatom clustering, one could first argue that the formation of chiral bulk currents around the thallium islands is, similarly to the situation of homogenous distribution, locally dominated by intrinsic spin-orbit coupling, with negligible impact of the Rashba SOC term. Nevertheless one cannot totally exclude a finite contribution of a Rashba SOC, which would limit the spin lifetime in the situation of SHE and anomalous metallic state discussed in the main manuscript.

As a lower bound however, by making a crude extrapolation from the spin relaxation time value obtained for gold adatom [3], one estimates that the spin diffusion length would be on the order of hundred nanometers, thus warranting the robustness of the obtained quantum phases for the considered system size. Calculations following the methodology implemented in [3] should further quantify the precise spin lifetime at play and experimental conditions for observation of reported multiple quantum phases.

## Local density of spin accumulation

Figure S1 shows the local spin density given by the difference between the local density of occupied states for spin-up and spin-down electrons, for the same systems and energies considered in Fig. 2 of the main text. The red (positive) regions indicate a predominant density of spin up, while the blue (negative) regions indicate a predominant density of spin down. In agreement with our simulations of the spectral currents, we find that the edges are fully polarized for islands with radius $r$=0.5 nm (panel (a)), roughly polarized for $r$=1.5 nm (panel (b)), and not well polarized for $r$=2 nm (panel (c)).

Note that, in the latter case the system still shows a large region with the same polarization, thus indicating that the formation of local chiral currents is still present, as confirmed by our 2D simulations. Accordingly, we can expect some spin accumulation at the edges for much larger ribbons and for larger islands, as experimentally observed in [4].

We summarize our results concerning the spin polarization of the edges by considering the index $y_{SHE} = \langle y^\uparrow \rangle - \langle y^\downarrow \rangle$, which quantifies the difference in the expected average transverse positions of the spin-up $\langle y^\uparrow \rangle$ and spin-down $\langle y^\downarrow \rangle$ electrons. If there is no spin accumulation, this quantity vanishes, while it is maximal in the case of the QSHE. Figure S2 shows how this quantity evolves as a function of island size at an energy E=-33.5 meV for the 50 nm wide ribbon considered in the paper.

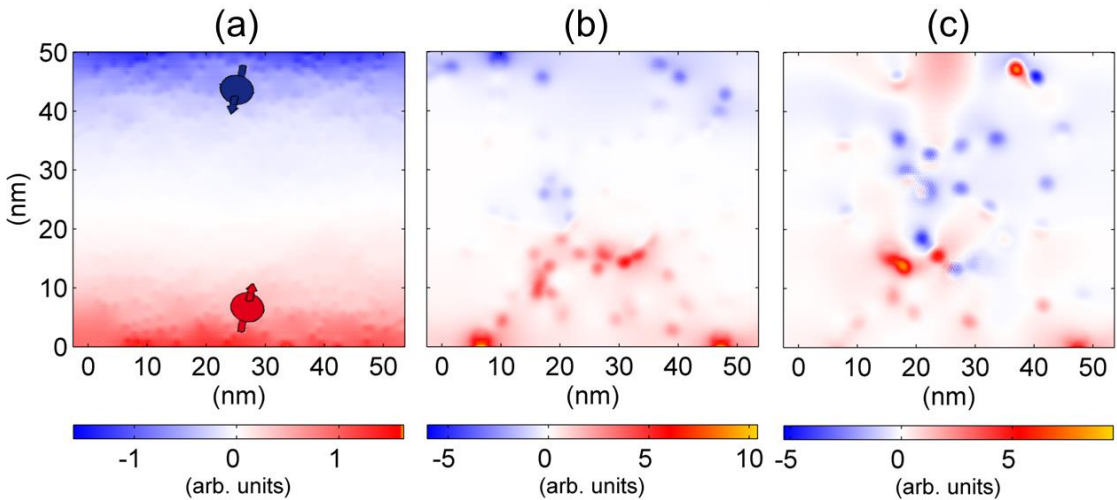

Fig.S1: Local spin polarization for a 50 nm wide armchair GNR with a 15% coverage of thallium adatoms over a section of 50 nm, segregated into islands with radius (a) $r$=0.5 nm (and energy -33.5 meV), (b) $r$=1.5 nm (and energy 21.5 meV), (c) $r$=2 nm (and energy -33.5 meV).

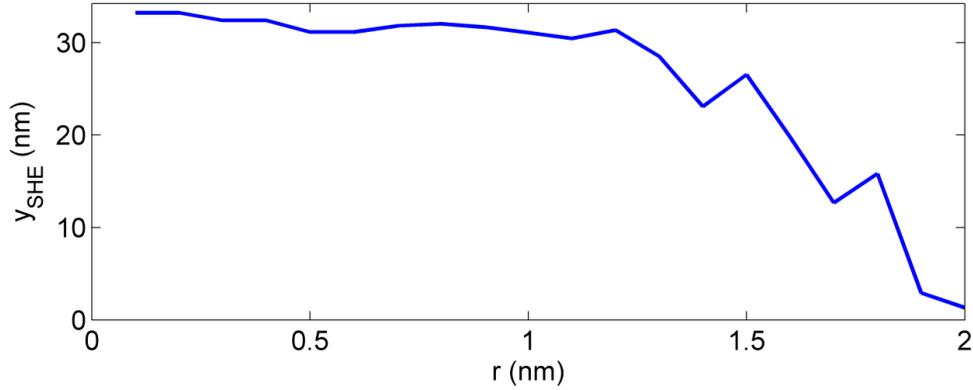

Fig.S2: Evolution of the $y_{SHE}$ for a 50 nm wide GNR with a 15% coverage of thallium adatoms over a section of 50 nm, segregated into islands with radius up to 2 nm at energy -33.5 meV.

### Intuitive cartoon picture of the crossover mechanism

Figures 3(c-e) of the paper suggest that chiral currents flow clockwise or counterclockwise around (and partially inside at least for the small radii we consider) the islands depending on the spin polarization of the electrons. This phenomenon is schematically pictured in Fig. S3(a). When two islands are close enough, the current flowing at their common surface cancels out. Therefore, we obtain a chiral current flowing through the group of islands. These chiral channels are the key ingredient to understand our main results.

When there is a high coverage of adatoms, this behavior leads to the formation of chiral edge channels, see Fig. S3(b), similarly to the situation in the ordinary quantum Hall effect. The currents are then driven to flow along the edges, as seen in our simulations.

When the islands are slightly larger, for given adatom concentration, the domain is partially fragmented in separate regions with high adatom coverage (the islands or groups of islands) and regions without adatoms (between the islands). As illustrated in Fig. S3(c), in addition to the edge channels we observe the rise of local chiral channels around isolated groups of islands and non-chiral currents (green zones) where the non-functionalized regions are large enough. This entails the presence of bulk currents that drive the conductance above the plateau.

Finally, for larger islands, adatoms are concentrated in clusters and relatively large areas of the ribbon are free of adatoms. In this case, the edge channels disappear (because of the low adatom coverage at the edges) and the chiral currents are concentrated in the islands. The regions between islands allow for non-chiral currents. The extension of the chiral vortices and their connection operated by the non-chiral currents allow the formation of extended bulk channels, which are robust against localization, as reported

in our simulations. An extremely large distance between the islands might lead, in the presence of disorder, to the localization of the electron in the non-functionalized areas, with the consequent quenching of the bulk metallic states.

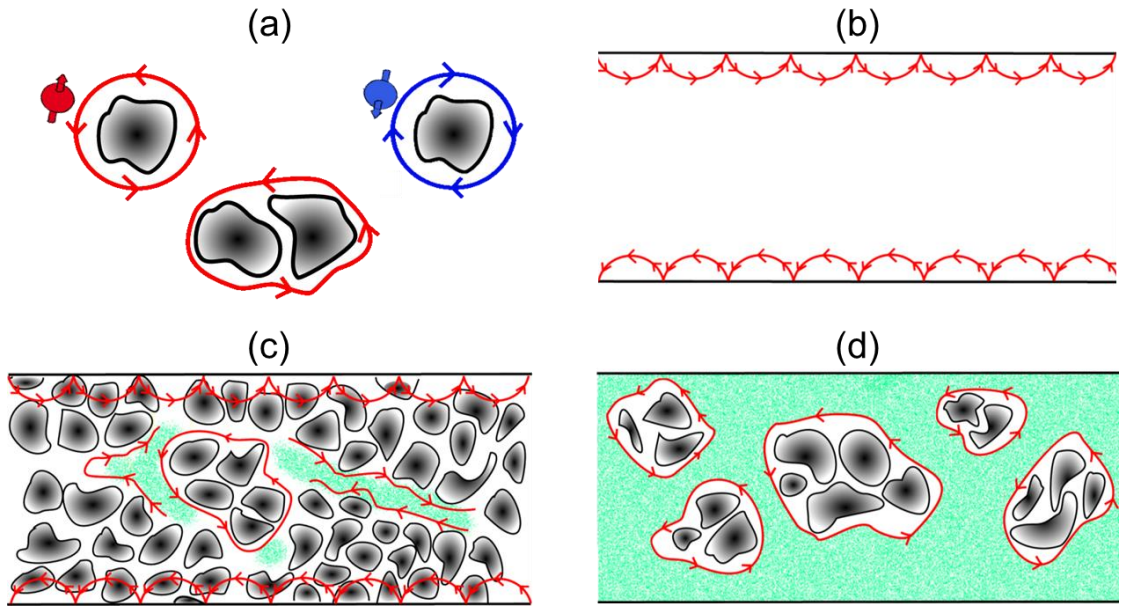

Fig.S3: (a) Local counter-clockwise and clockwise chiral currents around an island for spin-up (red) and spin-down (blue) electrons, respectively. Spin-up chiral current around two close islands. (b) Formation of the chiral edge states for high coverage of adatoms. (c) Presence of chiral vortices around isolated groups of islands, edge currents and non-chiral currents in adatom free area (green regions) in case of moderate coverage. (d) Chiral currents around the groups of islands and non-chiral bulk current (green region) in case of large islands. Edge states are suppressed.

Note that the local charge neutrality point under the islands is considerably shifted with respect to that of the adatom free areas. As a consequence, the current density and density of states is much higher in the region of the clusters than in between.

In conclusion, the proposed mechanism provides a general and intuitive explanation of what has been reported in both two-terminal and 2D simulations and for different size of the clusters. We can summarize the overall picture as follows:

- For small islands and high coverage of thallium adatoms, a topological phase with edge states develops in absence of any bulk current.

- In the situation of adatom segregation and island formation (with r below 2 nm), chiral currents emerge in the bulk with the tendency to produce spin accumulation at graphene edges.

- For larger islands ($r$ = 2 nm and more), large non-chiral regions coexist with some locally current chiral paths in the vicinity of islands. In two-terminal transport,

4

electrons penetrate deep in the bulk with only a remaining poor residual spin accumulation at the edges. In 2D transport, the chiral paths prevent electrons from localizing and preserve the diffusive nature of electronic transport.

- In the limit of large island radius, the system becomes highly inhomogeneous, with significantly large non chiral regions, which eventually dominate over local spin transport chirality produced close to the island. If additional disorder is active in between islands, localization effects are expected to develop.

**References**

[1] C. Weeks, J. Hu, J. Alicea, M. Franz, and R. Wu, Phys. Rev. X **1**, 021001 (2011).

[2] C. L. Kane and E. J. Mele, Phys. Rev. Lett. **95**, 146802 (2005).

[3] D. Van Tuan, F. Ortmann, D. Soriano, S. O. Valenzuela and S. Roche, Nature Phys. **10**, 857 (2014).

[4] J. Balakrishnan, G. K. W. Koon, A. Avsar, Y. Ho, J. H. Lee, M. Jaiswal, S.-J. Baeck, J.-H. Ahn, A. Ferreira, M. A. Cazalilla, A. H. Castro Neto, and B. Özyilmaz, Nat. Commun. **5**, 4748 (2014).




\begin{thebibliography}{48}
\bibitem [{not()}]{note}%
  \BibitemOpen
  \href@noop {} {\bibinfo  {journal} {Following \cite{KAN_PRL95b} and
  \cite{WEE_PRX1}, the Rashba SOC term has been neglected here, precluding spin
  population mixing. Relaxing such an approximation should yield finite spin
  lifetime, which will limit the system size where the coherent quantum
  phenomena occur \cite{SM}}\ }\BibitemShut {NoStop}%
\end{thebibliography}
\end{document}